\documentclass[english]{revtex4-1}
\usepackage[LGR,T1]{fontenc}
\usepackage[latin9]{inputenc}
\usepackage{geometry}
\usepackage{array}
\usepackage{float}
\usepackage{multirow}
\usepackage{amssymb}

\makeatletter

\DeclareRobustCommand{\greektext}{%
  \fontencoding{LGR}\selectfont\def\encodingdefault{LGR}}
\DeclareRobustCommand{\textgreek}[1]{\leavevmode{\greektext #1}}
\ProvideTextCommand{\~}{LGR}[1]{\char126#1}

\providecommand{\tabularnewline}{\\}

\makeatother

\usepackage{babel}
\begin{document}

\title{A Minimal Fermion-Scalar Preonic Model}

\author{U. Kaya}

\address{Department of Physics, Faculty of Sciences, Ankara University, Ankara,
Turkey}
\email{ukaya@etu.edu.tr}

\selectlanguage{english}%

\author{B. B. Oner}

\address{Department of Material Science and Nanotechnology, TOBB University
of Economics and Technology, Ankara, Turkey}
\email{b.oner@etu.edu.tr}

\selectlanguage{english}%

\author{S. Sultansoy}

\address{TOBB University of Economics and Technology, Ankara, Turkey}

\address{ANAS Institute of Physics, Baku, Azerbaijan}
\email{ssultansoy@etu.edu.tr}

\selectlanguage{english}%
\begin{abstract}
A minimal fermion-scalar preonic model containing two fermionic and
one scalar preons is proposed. This scheme allows to prevent the occurrence
of undesired SM-level particles, namely leptons and quarks with unusual
electric charges. Similar to the previous FS models, color-octet leptons
and color-sextet quarks, which are expected to have masses much lower
than the compositeness scale, are predicted. Observation of these
particles could provide first indications of preonic models. FCC/SppC
pp option will give opportunity to probe $m_{q_{6}}$ up to 48/75
TeV and $m_{l_{8}}$ up to 15/27 TeV within one year operation at
nominal luminosity. FCC/SppC based ep and $\mu$p colliders will essentially
enlarge covered mass region, namely, $m_{e_{8}}$ up to 23/27 TeV
and $m_{\mu_{8}}$ up to 68/80 TeV.
\end{abstract}
\maketitle

\section{Introduct\i on}

Structure of the atom has been revealed by the famous Rutherford experiment
which was performed almost a century ago \cite{key-1}. In 1930s,
the nucleus of the atom is discovered to be a bound state of protons
and neutrons. Thus, a scientific basis have been constructed for Periodical
Table of chemical elements. In 1960s, high energy physics experiments
showed that hadrons (including protons and neutrons) were also bound
states of more fundamental particles: quarks \cite{key-2}. Thanks
to these experiments, Standard Model (SM) was constituted which seems
to be in conformity with succeeding experiments in $\lesssim$ TeV
energy region \cite{key-3}. On the other hand, a lot of phenomena
(such as family replication, fermion masses and mixings, left-right
asymmetry etc.) still can not be explained by the SM. Several approaches
reaching beyond the Standard Model (BSM) have been proposed in order
to address these problems.

One of the promising branches of these BSM proposals is composite
models of quarks and leptons. Existence of at least three fermion
families and observation of the inter-family mixings of quarks and
leptons support the idea of the existence of a more fundamental level
of matter. Pati and Salam have denoted these fundamental particles
as preons. Historical arguments favoring preonic models are presented
in Table I \cite{key-5,key-6}. Composite models started to be developed
from 1970s (see \cite{key-7} and references therein) and can be divided
into two main subclasses: fermion-scalar (FS) and three fermion (FFF)
models. 

Even though there have not been any direct experimental evidence indicating
a substructure of the SM fermions yet, mass pattern of fermion families
and CKM mixings can be regarded as a manifestation of compositeness
of these fermions. Future highest energy colliders such as FCC \cite{key-8,key-9}
and/or SPPC \cite{key-10} with $\sqrt{s}$=100/136 TeV, which are
planned to be constructed in 2030s, will enable us to investigate
the new physics at the multi-TeV scale. Let us denote the new compositeness
scale as $\Lambda$. A comparison between $\Lambda$ and center of
mass energies, $\sqrt{s}$, of future colliders points out our expectations
from these colliders. If $\sqrt{s}\ll\Lambda$, compositeness induced
contact four fermion interactions of SM particles have usually been
considered, since one expects that the masses of new particles lay
in the order of $\Lambda$. If $\sqrt{s}>\Lambda$, interactions and
particles of the new physics are expected to be revealed and if this
scheme will be realized at future colliders, expected results of these
high energy collisions would vary by a selected preonic model heavily.
The compositeness scale of the new physics, $\Lambda$, is quietly
larger than the masses of SM fermions ($m_{SM}$). Currently there
are three known mechanisms to satisfy the condition $m_{SM}\ll\Lambda$:
chiral protection, quasi-Goldstone fermion mechanisms (for details
see \cite{key-7} and references therein), as well as flavor democracy
\cite{key-36,key-37} (which provides the opportunity to get the massless
states as the superposition of initially massive particles and, therefore,
gives opportunity to handle \textquotedbl{}massless\textquotedbl{}
composite objects within preonic models). The true protection mechanism,
either one of the abovementioned or currently unknown mechanism, will
be clarified after the discovery of preonic dynamics. 

Commonly, fermion-scalar models up to now include two fermionic and
two scalar preons. In this work, in a belief of minimality at the
ultimate fundamental physics scale, we show that it is possible to
set a more economic preonic model containing two fermions and one
scalar. In Section II conventional (2 fermion, 2 scalar) preon models
are given with a short summary. In Section III, preonic set of the
current study is presented. Afterwards, predicted SM-level exotic
particles are described in Section IV and finally conclusion/final
remarks are given in a short summary in Section V.
\begin{center}
\begin{table}[H]
\caption{Examples of \textquotedblleft fundamental\textquotedblright{} particle
inflations encountered in the last century. }

\centering{}%
\begin{tabular}{|c|c|c|c|}
\hline 
Stages & 1870s-1930s & 1950s-1970s & 1970s-2020s\tabularnewline
\hline 
\hline 
Fundamental Particle Inflation & Chemical Elements & Hadrons & Quarks \& Leptons\tabularnewline
\hline 
Systematic & Periodic Table & Eight-fold Way & Family Replication\tabularnewline
\hline 
Confirmed Predictions & New Elements & New Hadrons & $l_{8}$ and $q_{6}$?\tabularnewline
\hline 
Clarifying Experiments & Rutherford & SLAC DIS & LHC? or rather FCC?\tabularnewline
\hline 
Building Blocks & Proton, Neutron, Electron & Quarks & Preons?\tabularnewline
\hline 
Energy Scale & MeV & GeV & Multi-TeV?\tabularnewline
\hline 
Impact on Technology & Exceptional & Indirect & Exceptional?\tabularnewline
\hline 
\end{tabular}
\end{table}
\par\end{center}

\section{Fermion-Scalar Models}

Fermion-scalar (FS) type composite models were proposed forty years
ago \cite{key-1212,key-1313,key-1414}. Most of FS preonic models
assume existence of two fermionic and two scalar preons. Below we
assume that preons are color triplets. In this case color singlet
SM leptons are predicted to be bound states of one fermionic preon
and one scalar anti-preon

\begin{eqnarray}
l & = & (F\bar{S})=1\oplus8
\end{eqnarray}

$\vphantom{}$

\noindent with a color-octet partner $l_{8}$. Quarks are expected
to be composed of one fermionic and one scalar anti-preons in a similar
manner 

\begin{eqnarray}
q & = & (\bar{F}\bar{S})=3\oplus\bar{6}
\end{eqnarray}

$\vphantom{}$

\noindent which means that each SM quark has one anti-sextet partner
$\bar{q}_{6}$. 

The first SM family fermions are given as

\begin{eqnarray}
\nu_{e} & = & (F_{1}\bar{S_{1}})\qquad e=(F_{2}\bar{S_{1}})\qquad d=(\bar{F_{1}}\bar{S_{2}})\qquad u=(\bar{F_{2}}\bar{S_{2}}).
\end{eqnarray}

$\vphantom{}$

Table II represents possible electric charge set schemes under an
assumption $|Q_{F,S}\leq1|$ \cite{key-15}. The third column (Model
III) set of the table corresponds to Fritzsch-Mandelbaum model \cite{key-16}
and the option given in the fourth column (Model IV) implies the fermion-scalar
symmetry from electric charge viewpoint which may be indication of
super-symmetry at preonic level. 
\begin{center}
\begin{table}[H]
\caption{Electric charges of scalar and fermionic preons.}

\centering{}%
\begin{tabular}{|c|c|c|c|c|c|}
\hline 
\multirow{2}{*}{Preons} & \multicolumn{5}{c|}{Electric Charges}\tabularnewline
\cline{2-6} 
 & Model I & Model II & Model III & Model IV & Model V\tabularnewline
\hline 
\hline 
$F_{1}$ & 0 & 1/3 & 1/2 & 2/3 & 1\tabularnewline
\hline 
$S_{1}$ & 0 & 1/3 & 1/2 & 2/3 & 1\tabularnewline
\hline 
$F_{2}$ & -1 & -2/3 & -1/2 & -1/3 & 0\tabularnewline
\hline 
$S_{2}$ & 1/3 & 0 & -1/6 & -1/3 & -2/3\tabularnewline
\hline 
\end{tabular}
\end{table}
\par\end{center}

One of the main problems of conventional FS models is some undesirably
predicted SM-level particles which have not been observed yet. For
example, particles below are predicted in addition to the first SM
family fermions: color singlets
\begin{center}
$(F_{1}\bar{S_{2}})$ and $(F_{2}\bar{S_{2}})$
\par\end{center}

\noindent and color triplets
\begin{center}
$(\bar{F_{1}}\bar{S_{1}})$ and $(\bar{F_{2}}\bar{S_{1}})$.
\par\end{center}

\noindent Electric charges of these new particles are presented in
Table III.
\begin{center}
\begin{table}[H]
\caption{Electric charges of the additional undesired fermions corresponding
to the preonic sets given in Table II.}

\centering{}%
\begin{tabular}{|c|c|c|c|c|c|}
\hline 
\multirow{2}{*}{Additional Particles} & \multicolumn{5}{c|}{Electric Charges}\tabularnewline
\cline{2-6} 
 & Model I & Model II & Model III & Model IV & Model V\tabularnewline
\hline 
\hline 
$\begin{array}{c}
\\
(F_{1}\bar{S_{2}})\\
\\
\end{array}$ & -1/3 & 1/3 & 2/3 & 1 & 5/3\tabularnewline
\hline 
$\begin{array}{c}
\\
(F_{2}\bar{S_{2}})\\
\\
\end{array}$ & -4/3 & -2/3 & -1/3 & 0 & 2/3\tabularnewline
\hline 
$\begin{array}{c}
\\
(\bar{F_{1}}\bar{S_{1}})\\
\\
\end{array}$ & 0 & -2/3 & -1 & -4/3 & -2\tabularnewline
\hline 
$\begin{array}{c}
\\
(\bar{F_{2}}\bar{S_{1}})\\
\\
\end{array}$ & 1 & 1/3 & 0 & -1/3 & -1\tabularnewline
\hline 
\end{tabular}
\end{table}
\par\end{center}

There is no reason for these additional particles to be absent and
to have masses far above the SM scale. Fritzsch and Mandelbaum proposed
QED- or QCD-like preon dynamics (hypercolor) that resolves this problem
\cite{key-16}: repulsive interactions between preons with the same
hypercolor charges prevent these undesired bound states. However,
in their model, $S_{1}$ is color anti-triplet, whereas $F_{1}$,
$F_{2}$ and $S_{2}$ are color triplets. Moreover, preon dynamics
need not to be QED- or QCD-like. For example, \textquotedblleft gravitation-like\textquotedblright{}
dynamics involve attractive force only.

\section{A Minimal fermion-scalar model}

In this study, considering the problem above, we propose a novel minimal
FS model which prevents the occurence of undesired SM-level particles.
Proposed preons and their color, charge and spins are given in Table
IV. It should be noted that electric charge set is unique, while in
non-minimal models there are 5 choices (see Table II).
\begin{center}
\begin{table}[H]
\caption{Color, charge and spins of minimal FS model preons.}

\centering{}%
\begin{tabular}{|c|c|c|c|}
\cline{2-4} 
\multicolumn{1}{c|}{} & Color (C) & Charge (Q) & Spin (S)\tabularnewline
\hline 
$F_{1}$ & 3 & 1/6 & 1/2\tabularnewline
\hline 
$F_{2}$ & 3 & -5/6 & 1/2\tabularnewline
\hline 
$S$ & 3 & 1/6 & 0\tabularnewline
\hline 
\end{tabular}
\end{table}
\par\end{center}

\noindent In this case, bound states of fermionic preons with the
scalar preon constitute the first SM family fermions as below:

\[
Q_{F_{1}}+Q_{\bar{S}}=0,\quad C_{F_{1}}\otimes C_{\bar{S}}=3\otimes\bar{3}=1\oplus8\quad\rightarrow\quad\nu_{e}\equiv(F_{1}\bar{S})
\]

\[
Q_{F_{2}}+Q_{\bar{S}}=-1,\quad C_{F_{2}}\otimes C_{\bar{S}}=3\otimes\bar{3}=1\oplus8\quad\rightarrow\quad e\equiv(F_{2}\bar{S})
\]

\[
Q_{\bar{F}_{1}}+Q_{\bar{S}}=-1/3,\quad C_{\bar{F_{1}}}\otimes C_{\bar{S}}=\bar{3}\otimes\bar{3}=3\oplus\bar{6}\quad\rightarrow\quad d\equiv(\bar{F}_{1}\bar{S})
\]

\[
Q_{\bar{F}_{2}}+Q_{\bar{S}}=2/3,\quad C_{\bar{F_{2}}}\otimes C_{\bar{S}}=\bar{3}\otimes\bar{3}=3\oplus\bar{6}\quad\rightarrow\quad u\equiv(\bar{F}_{2}\bar{S})
\]

\noindent One should note that, model still predicts color octet leptons
and color sextet quarks. 

Preons in fermion-scalar models are color triplets which means QCD
is realised at preonic level. If space-time structure is not changed,
it is natural to assume that electro-weak gauge symmetry is also realised
at preonic level. We present weak iso-spin and weak hypercharge values
for preons in Table V for this reason.
\begin{center}
\begin{table}[H]
\caption{Weak iso-spin and weak hypercharge quantum numbers for preons corresponding
to Table IV regarding chirality of preonic level fermions.}

\centering{}%
\begin{tabular}{|c|c|c|}
\cline{2-3} 
\multicolumn{1}{c|}{} & Weak Isotopic Charge ($I_{3}$) & Weak Hypercharge ($Y$)\tabularnewline
\hline 
$\left(\begin{array}{c}
F_{1_{L}}\\
F_{2_{L}}
\end{array}\right)$ & $\begin{array}{c}
1/2\\
-1/2
\end{array}$ & -2/3\tabularnewline
\hline 
$F_{1_{R}}$ & 0 & 1/3\tabularnewline
\hline 
$F_{2_{R}}$ & 0 & -5/3\tabularnewline
\hline 
S & 0 & 1/3\tabularnewline
\hline 
\end{tabular}
\end{table}
\par\end{center}

Another important issue is related with family replication. As mentioned
in the Introduction, mass pattern of fermion families is another indication
of substructure(s) at a more fundamental level. The second and the
third SM fermion families can be constructed by quantum pair excitations
\cite{key-17}. For example, second family fermions may be constructed
by addition of $(S\bar{S})$ to first family fermions:
\begin{center}
$\nu_{\mu}\equiv(F_{1}\bar{S})(S\bar{S})$
\par\end{center}

\begin{center}
$\mu\equiv(F_{2}\bar{S})(S\bar{S})$
\par\end{center}

\begin{center}
$s\equiv(\bar{F}_{1}\bar{S})(S\bar{S})$
\par\end{center}

\begin{center}
$c\equiv(\bar{F}_{2}\bar{S})(S\bar{S})$
\par\end{center}

In a similar manner the third family can be expressed as:
\begin{center}
$\nu_{\tau}\equiv(F_{1}\bar{S})(S\bar{S})^{2}$
\par\end{center}

\begin{center}
$\tau\equiv(F_{2}\bar{S})(S\bar{S})^{2}$
\par\end{center}

\begin{center}
$b\equiv(\bar{F}_{1}\bar{S})(S\bar{S})^{2}$
\par\end{center}

\begin{center}
$t\equiv(\bar{F}_{2}\bar{S})(S\bar{S})^{2}$
\par\end{center}

In structures above we assume that only singlet component of $(S\bar{S})$
takes part in formation of the second and the third SM family fermions.
Alternatively, one can consider the case when color octet component
of $(S\bar{S})$ is also included in formation of the upper families.
In this case $(F\bar{S})(S\bar{S})$ has following color structure:
\begin{center}
$(1\oplus8)(1\oplus8)=1\oplus8\oplus8\oplus1\oplus8\oplus8\oplus10\oplus\overline{10}\oplus27$
\par\end{center}

\noindent Therefore, one can identify the first singlet as $\mu$
and the second singlet as $\tau$. As a result, muon has two color
octet partners, whereas $\tau$ lepton has two octet, one decuplet,
one anti-decuplet and one 27-plet partners. The same decomposition
takes place for $\nu_{\mu}$ and $\nu_{\tau}$.

\section{Color-octet Leptons and Color-sextet Quarks}

All of the preonic FS models predict color-octet leptons, $l_{8}$,
and color-sextet quarks, $q_{6}$. $SU_{W}(2)\times U_{Y}(1)$ structures
of $l_{8}$ and $\bar{q}_{6}$ are coincide with that of $l$ and
$q$, respectively. Therefore, chirality protection mechanism, which
keeps SM fermions' masses small, is also assumed to be valid for $l_{8}$
and $q_{6}$, such that $m_{l_{8}}$, $m_{q_{6}}\ll\Lambda$. Let
us mention that masses of the vector and scalar bound states (including
leptoquarks) are expected to be at the order of $\Lambda$. Therefore,
discovery of $l_{8}$ and $q_{6}$ at future high energy colliders
may provide a first confirmation of preonic models.

Production, signatures and discovery limits of color-sextet quarks
and color-octet leptons at the LHC have been roughly considered in
\cite{key-15}. In recent papers \cite{key-21,key-20,key-18,key-19}
$l_{8}$ production at the LHC have been analyzed in details: it is
shown that $m_{l_{8}}\lesssim1.2$ TeV is excluded by current ATLAS/CMS
data and future LHC runs will cover $m_{l_{8}}$ up to 2.5$\div$3
TeV. Certainly, future 100 TeV center of mass energy $pp$ colliders,
FCC and/or SppC, have a great potential for BSM physics search. In
Table VI we present discovery limits for resonant $q_{6}$ and pair
$l_{8}$ production at these colliders. 

Resonant $l_{8}$ production could be investigated at the FCC and
SppC based energy frontier $lp$ colliders (for main parameters of
FCC-$lp$ and SppC-$lp$ see \cite{key-22} and \cite{key-23}, respectively).
Potential of FCC-$ep$ for $e_{8}$ search has been analyzed in \cite{key-6},
similar analysis for $\mu_{8}$ at FCC-$\mu p$ have been performed
in \cite{key-24}. Discovery limits results are summarized in Table
VII and VIII, respectively.

\begin{table}[H]
\caption{Discovery ($5\sigma$) limits for $q_{6}$ and $l_{8}$ at future
$pp$ colliders.}

\centering{}%
\begin{tabular}{|c|c|c|c|c|}
\hline 
\multirow{2}{*}{Collider} & \multirow{2}{*}{$\sqrt{s}$, TeV} & \multirow{2}{*}{$L_{int}$ per year} & \multirow{2}{*}{$M_{l_{8}}$, TeV} & \multirow{2}{*}{$M_{q_{6}}$, TeV}\tabularnewline
 &  &  &  & \tabularnewline
\hline 
\hline 
LHC & 14 & 100 $fb^{-1}$  & 3 & 8\tabularnewline
\hline 
FCC & 100  & 500 $fb^{-1}$  & 15 & 48\tabularnewline
\hline 
SppC & 136  & 10000 $fb^{-1}$  & 27 & 75\tabularnewline
\hline 
\end{tabular}
\end{table}

\begin{table}[H]
\caption{Discovery ($5\sigma$) limits for $e_{8}$ at FCC/SppC based ($E_{p}=50/68$
TeV) $ep$ colliders.}

\centering{}%
\begin{tabular}{|c|c|c|c|}
\hline 
\multirow{2}{*}{$E_{e}$, GeV} & \multirow{2}{*}{$\sqrt{s}$, TeV} & \multirow{2}{*}{$L_{int}$ per year} & \multirow{2}{*}{$M_{e_{8}}$, TeV }\tabularnewline
 &  &  & \tabularnewline
\hline 
\hline 
60 & 3.46/4.04 & 100 $fb^{-1}$  & 2.9/3.3\tabularnewline
\hline 
\multirow{2}{*}{500} & \multirow{2}{*}{10.0/11.7} & 10 $fb^{-1}$  & 8.1/9.4 \tabularnewline
\cline{3-4} 
 &  & 100 $fb^{-1}$  & 8.6/10.0\tabularnewline
\hline 
\multirow{2}{*}{5000} & \multirow{2}{*}{31.6/36.9} & 1 $fb^{-1}$  & 20.1/23.4 \tabularnewline
\cline{3-4} 
 &  & 10 $fb^{-1}$  & 23.1/26.9\tabularnewline
\hline 
\end{tabular}
\end{table}

\begin{table}[H]
\caption{Discovery ($5\sigma$) limits for $\mu_{8}$ at FCC/SppC based ($E_{p}=50/68$
TeV) $\mu p$ colliders.}

\centering{}%
\begin{tabular}{|c|c|c|c|}
\hline 
\multirow{2}{*}{$E_{\mu}$, GeV} & \multirow{2}{*}{$\sqrt{s}$, TeV} & \multirow{2}{*}{$L_{int}$ per year} & \multirow{2}{*}{$M_{\mu_{8}}$, TeV }\tabularnewline
 &  &  & \tabularnewline
\hline 
\hline 
750 & 12.2/14.3 & 5/12 $fb^{-1}$  & 9.21/12.1\tabularnewline
\hline 
1500 & 17.3/20.2 & 5/43 $fb^{-1}$  & 13.2/20.2\tabularnewline
\hline 
20000 & 63.2/73.8 & 10 $fb^{-1}$  & 41.5/48.5\tabularnewline
\hline 
50000 & 100/117 & 10 $fb^{-1}$  & 68.4/80\tabularnewline
\hline 
\end{tabular}
\end{table}

\section{Conclusion}

In this study, we propose a novel fermion-scalar composite model to
form SM fermions from preonic level while assuming SM bosons as fundamental.
By means of the minimal approach of the model, fermion-scalar bound
states are constructed by only three preonic-level particles, namely
2 fermionic and 1 scalar preons. This scheme allows to prevent the
occurrence of undesired SM-level particles, namely leptons and quarks
with unusual electric charges. Similar to the previous FS models,
color-octet leptons and color-sextet quarks, which are expected to
have masses much lower than the compositeness scale, are predicted.
Observation of these particles could provide first indications of
preonic models. FCC (SppC) $pp$ option will provide opportunity to
probe $m_{q_{6}}$ up to 48 (75) TeV and $m_{l_{8}}$ up to 15 (27)
TeV within one year operation at nominal luminosity. FCC/SppC based
$ep$ and $\mu p$ colliders will essentially enlarge covered mass
region, namely, $m_{e_{8}}$ up to 23/27 TeV and $m_{\mu_{8}}$ up
to 68/80 TeV.
\begin{acknowledgments}
This study is supported by TUBITAK under the grant no 114F337.
\end{acknowledgments}

\end{document}